\documentclass[runningheads,a4paper]{llncs}

\usepackage{amssymb}
\usepackage{graphicx}

\begin{document}

\mainmatter  % start of an individual contribution

\title{Systematics on ground-state energies of nuclei within the neural networks}

\titlerunning{Systematics on ground-state energies of nuclei within the neural networks}

\author{Tuncay Bayram, Serkan Akkoyun and S. Okan Kara
}
\authorrunning{Tuncay Bayram, Serkan Akkoyun and S. Okan Kara}

\institute{Department of Physics, Sinop University, 57000 Sinop, Turkey\\
Department of Physics, Cumhuriyet University, 58140 Sivas, Turkey\\
Department of Physics, Nigde University, Nigde, Turkey\\
}

\toctitle{Systematics on ground-state energies of nuclei} \tocauthor{T. Bayram
et al.}
\maketitle

\begin{abstract}
One of the fundamental ground-state properties of nuclei is binding
energy. In this study, we have employed artificial neural networks
(ANNs) to obtain binding
energies based on the data calculated from Hartree-Fock-Bogolibov (HFB) method with
the two SLy4 and SKP Skyrme forces. Also, ANNs have been employed to obtain
two-neutron and two-proton separation energies of nuclei.
Statistical modeling of nuclear data using ANNs has been seen as to
be successful in this study. Such a statistical model can be
possible tool for searching in systematics of nuclei beyond existing
experimental nuclear data.

\keywords{Ground-state energies, artificial neural network, Hartree-Fock-Bogoliubov Method}
\end{abstract}

\section{Introduction}

The latest advances in computational physics and experimental
techniques have mock up renewed interest in nuclear structure
theory.  The development of the “universal” nuclear energy
density functional still remains one of the major challenges for
nuclear theory. While HFB methods have already achieved a level of
sophistication and precision which allows analyses of experimental
data for a wide range of properties and for arbitrarily heavy
nuclei~\cite{Stoitsov05}. Much work remains to be done. There are
many open questions when one moves towards the neutron and proton
drip lines and superheavy region, while behaviour of the region near
valley of stability is well understood. Developing a universal
nuclear density functional will require a better understanding of
the density dependence, isospin effects, and pairing, as well as an
improved treatment of symmetry-breaking effects and many-body
correlations.

For investigation of ground-state properties of nuclei in nuclidic
chart, there are three most prominent methods the Skyrme energy
functional, the Gogny model and relativistic mean field (RMF)
model~\cite{Ring80,Greiner96}. They have been formulated in terms of
effective density-dependent nucleon-nucleon interactions. The Skyrme
forces as an effective interaction are usually used within the fully
microscopic self-consistent mean field theories, because the
analytical simplicity of these forces provide easy determination of
the parameters from same basic properties of nuclei (e.g., the
saturation of infinite nuclear matter and binding energies of some
nuclei). However, their accuracy on the predictions for properties
of nuclei is related with the parameters. There is a number of
Skyrme forces in literature (details can be found in
Ref.~\cite{Chabanat98} and references therein). Three Skyrme forces
SKM*, SKP and SLy4 are widely used in calculations of the
ground-state properties of nuclei within the (HFB) method. In recent
decade,  results of large-scale ground-state calculations
($\sim$1500 nuclei) have been presented within the framework of HFB
method by using the SKP and SLy4 parameters~\cite{Dobaczewski}. In
the study of Ref.\cite{Dobaczewski}, the authors used the code
HFBTHO~\cite{Stoitsov05}. In the code HFB equations can be solved by
using the either axially deformed harmonic oscillator and
transformed harmonic oscillator basis.

The correct predictions of properties of nuclei play an essential
role for further mapping of nuclei in nuclidic chart. In
Ref.~\cite{Dobaczewski}, the authors have calculated ground-states
of nuclei such as binding energies, nuclear radii and quadrupole
moments. The binding energies of nuclei is one of the most
fundamental properties as well as nuclear size. The correct
prediction of the binding energy is important, because obtaining
theoretical predictions of two-nucleon separation energies and
alpha-decay properties of nuclei can be obtained from the calculated
binding energies.

In recent years, ANNs have been used in many fields in nuclear
physics as in the scientific areas, such as developing nuclear mass
systematics~\cite{Athanas2}, identification of impact parameter in
heavy-ion collisions~\cite{David,Bass,Haddad}, estimating beta decay
half-lives~\cite{Costiris}, investigating two photon exchange
effect~\cite{Graczyk} and obtaining nuclear charge
radii~\cite{Akkoyun}. In the present work, we used feed-forward
artificial neural networks (ANNs) in order to obtain binding and
two-nucleon separation energies of Sr, Xe, Er and Pb isotopes. The
fundamental task of the ANNs which have time advantage is to give
outputs through computation on the inputs. The method does not need
any relationship between input and output data. The main purpose of
the present study is showing of the ANNs successes in describing of
the unknown ground-state energies of nuclei by using known data. As
can be seen in this study, by using known data as an input, ANN
method can reproduce binding energies, two-neutron, two-proton
separation energies of unknown nuclei as consistent with theoretical
results.

The letter is organized as follows. In Section 2, the theoretical
framework for HFB and ANNs are given briefly. The results of this
study and discussions are presented in Section 3. Finally,
conclusions are given in Section 4.

\section{Artificial Neural Networks (ANNs)}

\begin{figure}
\centering
\includegraphics[height=5cm]{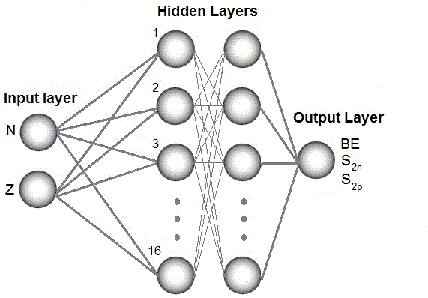}
\caption{2-16-16-1 ANN architecture that was used in this work.}
\label{fig1}
\end{figure}

Artificial neural networks (ANNs) \cite{Haykin} are mathematical
models that mimic the brain functionality. They are composed of
several processing units which are called neurons and they are
connected each other via adaptive synaptic weights. By this synaptic
connections, the neurons in the different layers communicate each
other and the data is transmitted. In our calculation, we have used
feed-forward ANN with four layers for estimating binding energies
and two nucleon separation energies. The first layer called input
layer consist of two neurons (corresponding to N and Z number of the
isotopes), the two intermediate layers named hidden layers are
composed of 16 neurons in each and the last one is output layer with
one neuron corresponding to the binding energy, neutron or proton
separation energies. The used architecture of the ANN was 2-16-16-1
(Fig.~\ref{fig1}) and the total numbers of adjustable weights were
304. No bias was used. The input neurons collect data from the
outside and the output neurons give the results. The hidden neuron
activation function was tangent hyperbolic
($tanh=(e^{x}-e^{-x})/(e^{x}+e^{-x})$) which is sigmoidlike
function. For details we refer the reader to Ref. \cite{Haykin}.

The ANN method would be a perfect tool to determine the binding and
two nucleon separation energies. One has to train network by known
isotopes data and then to feed the trained network with unknown
isotopes data in order to obtain neural network outputs for binding
and separation energies. In the learning stage of this work, a
back-propagation algorithm with Levenberg-Marquardt
\cite{Levenberg,Marquardt} were used for the change of the
connections between neurons in order to obtain aggrement between
neural network output and known output. The purpose of this training
stage is to minimize the difference between the outputs by
convenient modifications of synaptic connections. The error function
which measures this difference was mean square error (MSE). In the
training stage, $~$1300 isotopes except Sr, Xe, Er and Pb were used
for training of the ANN. The test of the trained ANN was performed on Sr, Xe, Er and Pb isotopes which have been never seen before by the network.

\section{Results and Discussions}

\begin{figure}
\centering
\includegraphics[height=5.5cm]{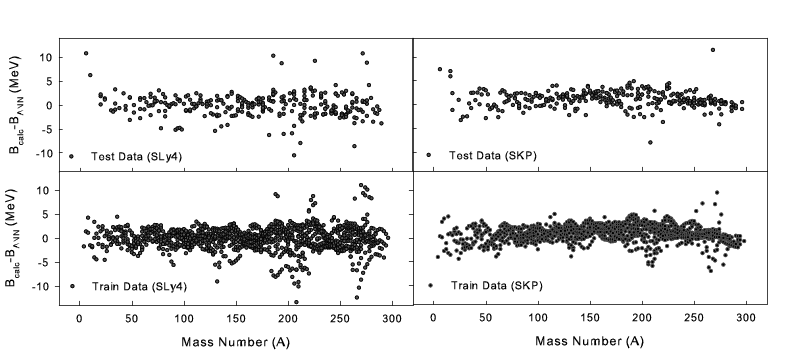}
\caption{The differences of total binding energy of nuclei between
the HFB calculations and ANN calculations for SLy4 (left panels) and
SKP (right panels) Skyrme parameters. Upper panels show the
differences for test data while lower panels show train data.}
\label{fig2}
\end{figure}

The ANNs, as in this work, are capable of learning systematics of
the binding and two nucleon separation energies for the isotopes
with high accuracy. In this study for different h numbers, the
minimum MSE values were between 0.00002 and 0.01 for training stage
and 0.00003 and 0.005 for testing stage. The differences in total
binding energy of nuclei between the HFB calculations and ANNs
results for SLy4 and SKP Skyrme parameters are shown in
Fig.~\ref{fig2}. As can be seen in the figure, the ANN outputs are
consistent with the calculations. The differences between calculated
results obtained from HFB method with SKP parameters and ANN outputs
are smaller than those of SLy4 Skyrme parameters.

\begin{figure}
\centering
\includegraphics[height=7cm]{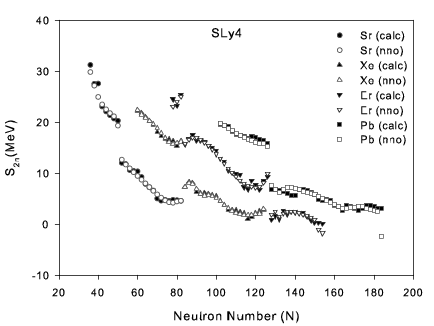}
\caption{Calculated two-neutron separation energies of Sr, Xe, Er
and Pb isotopic chains in HFB method with Skyrme force
SLy4~\cite{Dobaczewski} and the results of the present study with
ANN method} \label{fig3}
\end{figure}

\begin{figure}
\centering
\includegraphics[height=7cm]{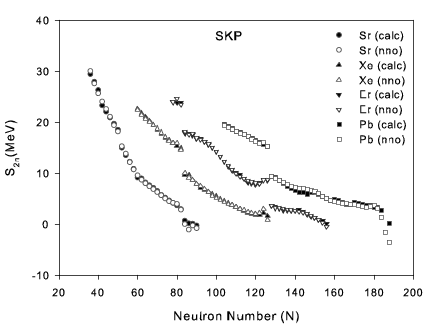}
\caption{Calculated two-neutron separation energies of Sr, Xe, Er
and Pb isotopic chains in HFB method with Skyrme force
SKP~\cite{Dobaczewski} and the results of the present study with ANN
method} \label{fig4}
\end{figure}

In Fig.~\ref{fig3} and~\ref{fig4}, we show two-neutron separation
energies of Sr, Xe, Er and Pb isotopic chains obtained by ANN method
based on HFB results with Skyrme forces SLy4 and SKP, respectively.
Also, the results of the HFB calculations taken
from~\cite{Dobaczewski} are shown in the same figures for
comparison. As can be seen in these figures, the predictions of
two-neutron separation energies of ANNs close to the input data.
This results show the predictive power of ANN method in the present
study. As is well known nuclei have a shell closure at magic neutron
numbers ($N=50$, 82 and 126). In particular, the correct predictions
of abrupt decreases of two-neutron separation energies at magic
neutron numbers are in agreement with the HFB predictions.

\begin{figure}
\centering
\includegraphics[height=7cm]{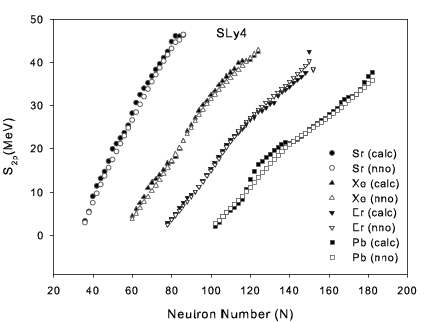}
\caption{Calculated two-proton separation energies of Sr, Xe, Er and
Pb isotopic chains in HFB method with Skyrme force
SLy4~\cite{Dobaczewski} and the results of the present study with
ANN method} \label{fig5}
\end{figure}

\begin{figure}
\centering
\includegraphics[height=7cm]{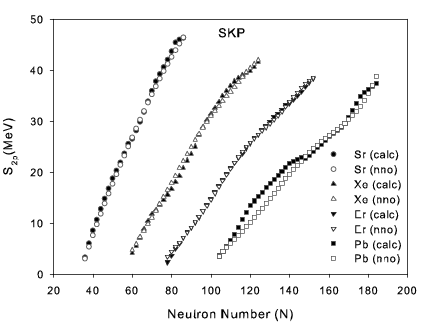}
\caption{Calculated two-proton separation energies of Sr, Xe, Er and
Pb isotopic chains in HFB method with Skyrme force
SKP~\cite{Dobaczewski} and the results of the present study with ANN
method} \label{fig6}
\end{figure}

 We have also obtained two-proton separation energies of Sr, Xe, Er and Pb
 isotopic chains with ANN method. The results are shown in Fig.~\ref{fig5}
 and~\ref{fig6} for SLy4 and SKP Skyrme forces, respectively. In the calculations
 same methodology has been followed as in the calculation of the two-neutron
 separation energies.  The results of the ANNs are in good agrement with the
 HFB calculations with Skyrme forces SLy4 and SKP, respectively. Also, the
 results of the HFB calculations taken from~\cite{Dobaczewski} are shown for
 comparison in the same figures.

\section{Summary}

Artificial neural network (ANN) has been employed to investigate the
ground-state energies of nuclei. In the calculations, the binding
energies of even-even 46 isotopic chains calculated in HFB method
with SLy4 and SKP Skyrme forces have been used as input data. The
ANN results for the ground-state binding energies of nuclei have
been obtained as in agreement with the original data. Also, we have
employed ANN method to obtain two-neutron and two-proton separation
energies of Sr, Xe, Er and Pb isotopic chains. The results show that
ANNs are capable in describing systematics of the ground-state
energies of nuclei. Particularly it has been understood that one
should take into account of predictive power of ANNs when dealing
with area which is unknown experimentally.


\begin{thebibliography}{99}
\bibitem{Stoitsov05} Stoitsov M V, Dobaczewski J
Nazarewicz W and Ring P 2005 Comp. Phys.
Commun. 167, 43.
\bibitem{Ring80} Ring P and Schuck P, 1980 The Nuclear
Many-Body Problem. Springer-Verlag, Berlin.
\bibitem{Greiner96} Greiner W and Maruhn J,
1996 Nuclear Models. Springer-Verlag, New York.
\bibitem{Chabanat98} Chabanat E, Bonche P,
Haensel P, Meyer J and Schaeffer, R 1998 Nucl. Phys. A 635, 231.
\bibitem{Dobaczewski} Skyrme-HFB deformed nuclear mass table, \url{http://www.fuw.edu.pl/~dobaczew/\\thodri/thodri.html}
\bibitem{Athanas2} Athanassopoulos S,
Mavrommatis E, Gernoth K A and Clark J W 2004 Nuclear Physics A 743, 222.
\bibitem{David} David C, Freslier M and Aichelin J
1995 Phys. Rev. C 51, 3, 1453.
\bibitem{Bass} Bass S A, Bischoff A, Maruhn J A
St\"{o}cker H and Greiner W 1996 Phys. Rev. C 53, 5, 2358.
\bibitem{Haddad} Haddad F, Hagel K, Li J,
Mdeiwayeh N, Natowitz J B, Wada R, Xiao B, David C,
Freslier M and Aichelin J 1997 Phys. Rev. C 55, 3,
1371.
\bibitem{Costiris} Costiris N, Mavrommatis
E, Gernoth K A and Clark J W 2007 arXiv:nucl-th/0701096v1.
\bibitem{Graczyk} Graczyk K.M 2011 Phys. Rev. C 84, 034314.
\bibitem{Akkoyun} Akkoyun S, Bayram T, Kara S O and
Sinan A 2012 arXiv:1212.6319 [nucl-th].
\bibitem{Haykin} Haykin S, 1999 Neural Networks: A Comprehensive
Foundation. Prentice-Hall Inc., Englewood Cliffs, NJ, USA.
\bibitem{Levenberg} Levenberg K, 1944 Quart. Appl. Math., Vol. 2, 164.
\bibitem{Marquardt} Marquardt D, 1963. SIAM J. Appl.
Math., Vol. 11, 431.
\end{thebibliography}
\end{document}